\documentclass[amsmath,amssymb,aip,apl,reprint,superscriptaddress,showpacs]{revtex4-1}

\usepackage{graphicx}
\usepackage{dcolumn}
\usepackage{bm}

\begin{document}

\title{Quantum spin Hall effect induced by electric field in silicene}

\author{Xing-Tao An}
\affiliation{School of Sciences, Hebei University of
Science and Technology, Shijiazhuang, Hebei 050018, China}
\affiliation{SKLSM, Institute of
Semiconductors, Chinese Academy of Sciences, P. O. Box 912, Beijing
100083, China}
\author{Yan-Yang Zhang}
\email{yanyang@semi.ac.cn} \affiliation{SKLSM, Institute of Semiconductors, Chinese Academy of
Sciences, P. O. Box 912, Beijing 100083, China}
\affiliation{International Center for Quantum Materials, Peking
University, Beijing 100871, China}
\author{Jian-Jun Liu}
\affiliation{Physics Department, Shijiazhuang University,
Shijiazhuang 050035, China}
\author{Shu-Shen Li}
\affiliation{SKLSM, Institute of Semiconductors, Chinese Academy of
Sciences, P. O. Box 912, Beijing 100083, China}

\date{\today}

\begin{abstract}
We investigate the transport properties in a zigzag silicene
nanoribbon in the presence of an external electric field. The staggered sublattice
potential and two kinds of Rashba spin-orbit couplings can be induced by the external electric field due to the
buckled structure of the silicene. A bulk gap is opened by the
staggered potential and gapless edge states appear in the gap by
tuning the two kinds of Rashba spin-orbit couplings properly. Furthermore, the gapless
edge states are spin-filtered and are insensitive to the non-magnetic disorder. These
results prove that the quantum spin Hall effect can be induced by an
external electric field in silicene, which may have certain practical significance
in applications for future spintronics device.
\end{abstract}

\maketitle

Recently, the quantum spin Hall effect (QSHE) has attracted
significant interests in the fields of condensed matter physics and
material science as it constitutes a new phase of matter and has
potential applications in spintronics.\cite{Kane, Bernevig, Konig,
CXLiu, CCLiu} The novel electronic state with a nontrivial
topological property and time-reversal invariance has a bulk energy
gap separating the valence and conduction bands and a pair of
gapless spin-filtered edge states sat the sample boundaries. The
QSHE has been first predicted by Kane and Mele in graphene in which
the intrinsic spin-orbit coupling opens a band gap at the Dirac
points.\cite{Kane} However, the QSHE can occur in graphene only at
unrealistically low temperatures since the intrinsic spin-orbit
coupling in graphene is rather weak.\cite{Yao, DHH, Min} Therefore,
it is crucial to search new materials with strong spin-orbit
coupling for realizing the QSHE. Recent theories and experiments
provide evidences of the QSHE in two-dimensional HgTe-CdTe quantum
wells.\cite{Konig, Bernevig}

Very recently, a close relative of graphene, a slightly buckled
honeycomb lattice of Si atoms called silicene has been
synthesized.\cite{Lalmi, Feng, Chen, Aufray, Padova} Silicene can be well compatible
with current silicon based electronic technology.
Many progresses in the study of silicene
have been made, both experimentally and theoretically. For example,
electronic properties and the giant magnetoresistance in silicene have
been reported.\cite{Ding1, Houssa, Ding2, Kang}
Moreover, almost every striking property of graphene
could be transferred to silicene.\cite{Padova, Houssa2}
It has been
theoretically shown that the strong intrinsic spin-orbit
coupling in silicene may lead to detectable QSHE.\cite{CCLiu,
Cahangirov, CCLiu2, Ezawa1, Ezawa2, An}

In this paper, we provide systematic investigations on the band
structures and electron transport properties of silicene in the
presence of an external electric field. Silicene consists of a
buckled honeycomb lattice of silicon atoms with two sublattices A
and B. We take a silicene sheet on the $x-y$ plane, and apply the
electric field in $z$ direction. The electric field generates a
staggered sublattice potential between silicon atoms at A sites and
B sites due to the buckled structure of the silicene. On the other
hand, two kinds of Rashba spin-orbit coupling, referring to the
nearest and next-nearest neighbor hoppings respectively, can also be
tuned by the external electric field. We find that a gap can be
opened by the staggered sublattice potential and gapless edge states
are induced in the gap by Rashba spin-orbit coupling. We predict
that the QSHE can be observed by applying an external electric field
in silicene even if the intrinsic spin-orbit coupling in the system
is very weak.

In the tight-binding representation, the silicene sample with an
external electric field can be described by the the following
Hamiltonian:\cite{CCLiu2}
\begin{eqnarray}
H&=&-t\sum_{\langle{ij}\rangle\alpha}c_{i\alpha}^{\dag}c_{j\alpha}+\sum_{i\alpha}\varepsilon_{i}\mu_{ij}c_{i\alpha}^{\dag}c_{i\alpha}\nonumber\\
&+&i\lambda_{R1}\sum_{\langle{ij}\rangle\alpha\beta}c_{i\alpha}^{\dag}(\vec{\sigma}\times\vec{d}_{ij}^0)^{z}_{\alpha\beta}c_{j\beta}\nonumber\\
&-&i\frac{2}{3}\lambda_{R2}\sum_{\langle\langle{ij}\rangle\rangle\alpha\beta}\mu_{ij}c_{i\alpha}^{\dag}(\vec{\sigma}\times\vec{d}_{ij}^0)^{z}_{\alpha\beta}c_{j\beta},
\end{eqnarray}
where $c_{i\alpha}^{\dag}$ creates an electron with spin
polarization $\alpha$ at site $i$; $\langle{ij}\rangle$ and
$\langle\langle{ij}\rangle\rangle$ run over all the nearest and
next-nearest neighbor hopping sites, respectively. The first term is
the nearest-neighbor hopping with the transfer energy $t=1.6eV$. The
second term is the staggered sublattice potential term, where
$\mu_{ij}=\pm1$ for the A (B) site and $\varepsilon_{i}$ is the
potential energy induced by the external electric field. The third
and fourth terms, respectively, represent the first Rashba
spin-orbit coupling associated with the nearest neighbor hopping and
the second Rashba spin-orbit coupling associated with the
next-nearest neighbor hopping. Both of them are induced by the
external electric field. Here $\vec{\sigma}=(\sigma_{x}, \sigma_{y},
\sigma_{z})$ is the Pauli matrix of spin and
$\vec{d}^{0}_{ij}=\vec{d}_{ij}/|\vec{d}_{ij}|$ with the vector
$\vec{d}_{ij}$ connecting two sites $i$ and $j$. The intrinsic SOC
term has been ignored intentionally since the main focus of this
work is the Rashba terms and the staggered potential, which
can be tuned by the external electric field.

We assume that the temperature is set to zero and two semi-infinite
silicene ribbons are employed as left and right leads. The
two-terminal conductance of the system can be calculated by the
nonequilibrium Green's function method and Landauer-B\"{u}ttiker
formula as
\begin{eqnarray}
G(E)=\frac{e^{2}}{h}\mathrm{Tr}[\mathbf{\Gamma}_{L}(E)\textbf{G}^{r}(E)\mathbf{\Gamma}_{R}(E)\textbf{G}^{a}(E)],
\end{eqnarray}
where
$\mathbf{\Gamma}_{p}(E)=i[\mathbf{\Sigma}_{p}^{r}(E)-\mathbf{\Sigma}_{p}^{a}(E)]$
is the line-width function and
$\textbf{G}^{r}(E)=[\textbf{G}^{a}(E)]^{\dag}=1/[\mathbf{E}-\textbf{H}_{cen}-\mathbf{\Sigma}_{L}^{r}-\mathbf{\Sigma}_{R}^{r}]$
is the retarded Green function with the Hamiltonian in the center
region $\textbf{H}_{cen}$.\cite{Ren} The self-energy
$\mathbf{\Sigma}_{p}^{r}$ due to the semi-infinite lead-$p$ can be
calculated numerically.\cite{Sancho}

With the help of the nonequilibrium Green's function method, the
local current flowing on site $i$ with spin $\sigma$ can be
expressed as
\begin{eqnarray}
J_{i\sigma}&=&\sum_{j\sigma'}J_{i\sigma,j\sigma'}=-\frac{e}{\hbar}\sum_{j\sigma'}[H_{i\sigma,j\sigma'}G^{<}_{j\sigma',i\sigma}(t,t)\nonumber\\
&-&G^{<}_{i\sigma,j\sigma'}(t,t)H_{j\sigma',i\sigma}],
\end{eqnarray}
where
$G^{<}_{i\sigma,j\sigma'}=i\langle{c^{\dag}_{j\sigma'}c_{i\sigma}}\rangle$
is the matrix element of the lesser Green¡¯s function of the
scattering region and $J_{i\sigma,j\sigma'}$ is the current from
site $i$ to $j$. After taking Fourier transformation, the local
current $J_{i\sigma,j\sigma'}$ can be calculated as
\begin{eqnarray}
J_{i\sigma,j\sigma'}=-\frac{2e}{\hbar}\int\frac{dE}{2\pi}\mathrm{Re}[H_{i\sigma,j\sigma'}G^{<}_{j\sigma',i\sigma}(E)].
\end{eqnarray}
Eq. (4) has been widely used in the local-current studies of
tight-binding models. \cite{Nikolic, Jiang, Xing} When the sample is
at zero temperature and the applied voltage is small, by applying
the Keldysh equation
$\mathrm{G}^{<}=\mathrm{G}^{r}(i\mathrm{\Gamma}_{L}f_{L}+i\mathrm{\Gamma}_{R}f_{R})\mathrm{G}^{a}$
with the Fermi distribution function $f_{p}=f_{0}(E+eV_{p})$,
\cite{Jauho} the Eq. (4) can be written as
\begin{eqnarray}
J_{i\sigma,j\sigma'}&=&\frac{2e}{h}\int^{eV_{R}}_{-\infty}dE\mathrm{Im}\{H_{i\sigma,j\sigma'}[G^{r}(\Gamma_{L}+\Gamma_{R})G^{a}]_{j\sigma',i\sigma}\}\nonumber\\
&+&\frac{2e^{2}}{h}(V_{L}-V_{R})\mathrm{Im}[H_{i\sigma,j\sigma'}G^{n}_{j\sigma',i\sigma}(E)],
\end{eqnarray}
where $V_{L}$ and $V_{R}$ are the voltages at the Lead-L and R,
respectively.
$\mathrm{G}^{n}(E)=\mathrm{G}^{r}(E)\mathrm{\Gamma}_{L}(E)\mathrm{G}^{a}(E)$
is electron correlation function. The first part of Eq. (5) can only
generate the equilibrium current and does not contribute to the
transport, so it can be dropped out in present work. It is the
second part that gives rise to the nonequilibrium current.

\begin{figure}[htb]
\centering
\includegraphics[scale=0.45,angle=0]{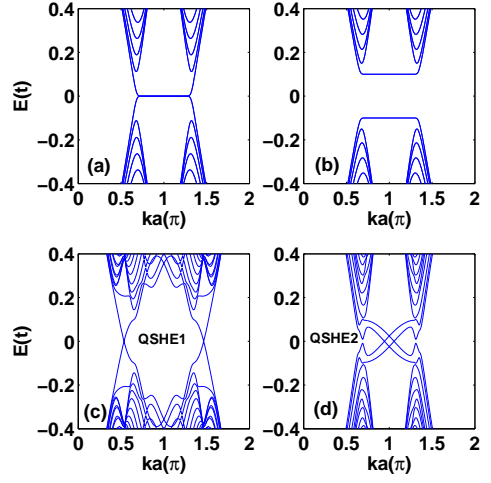}
\caption{Calculated energy bands in the zigzag
nanoribbon geometry for (a) $\varepsilon_{i}=0$, $\lambda_{R1}=0$
and $\lambda_{R2}=0$, (b) $\varepsilon_{i}=0.1t$, $\lambda_{R1}=0$
and $\lambda_{R2}=0$, (c) $\varepsilon_{i}=0.1t$,
$\lambda_{R1}=1.0t$ and $\lambda_{R2}=0.2t$, and (d)
$\varepsilon_{i}=0.1t$, $\lambda_{R1}=0.05t$ and
$\lambda_{R2}=0.6t$.} \label{figone}
\end{figure}

In the following numerical calculations, we use the hopping energy
$t$ as the energy unit. The width of the zigzag ribbon is $59a$,
where $a$ is the silicon-silicon distance. In Fig. \ref{figone} we
show the energy bands obtained from diagonalizing the tight-binding
Hamiltonian (1) with various parameters for a zigzag nanoribbon. The
nanoribbon with only the nearest-neighbor hopping shows a
semi-metallic behavior, as shown in Fig. \ref{figone} (a). An
energy gap can be opened due to the inversion symmetry breaking
induced by the staggered sublattice potential and the magnitude of
the gap is $2\varepsilon_{i}$ (see Fig. \ref{figone} (b)). When
the first and second Rashba spin-orbit couplings induced by the
external electric field are taken into account properly, which turn
silicene from normal insulating to quantum spin Hall regime, gapless
edge states appear within the band gap (see Figs. \ref{figone} (c)
and (d)). The gapless edge states with different spins connect the
conduction band and valence band. As usual, these gapless edge
states are originated from the nontrivial topological orders in the
bulk. According to different values of the first Rashba spin-orbit
coupling, the QSHE induced by the external electric field can be
divided into two types, QSHE1 and QSHE2. In QSHE1, when the first
Rashba spin-orbit coupling is strong, the edge states traverse the
bulk gap \emph{within} each valley, as shown in Fig. \ref{figone}
(c). However, in QSHE2 when the first Rashba spin-orbit coupling is
weak, the edge states inter-connect two valleys. Moreover, they bend
and give rise to ``subgaps'' around $K$ and $K'$, which makes the
structure of propagating channels complicated in the bulk gap.

\begin{figure}[htb]
\includegraphics[scale=0.45,angle=0]{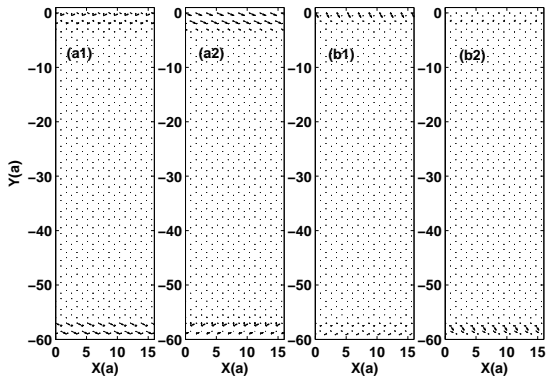}
\centering \caption{The local spin-up (a1 and b1) and
spin-down (a2 and b2) currents density vector distribution for (a)
$\varepsilon_{i}=0.1t$, $\lambda_{R1}=1.0t$ and $\lambda_{R2}=0.2t$
(QSHE1) with $E=0.05t$, and for (b) $\varepsilon_{i}=0.1t$,
$\lambda_{R1}=0.05t$ and $\lambda_{R2}=0.6t$ (QSHE2) with
$E=0.005t$.} \label{figtwo}
\end{figure}

To investigate the QSHE induced by the external electric field in
more details, the configurations of the spin-dependent
local-current-flow vector are plotted in Fig. \ref{figtwo}. We
focus only on the left-injected current. The Fermi energy is set to
be $E=0.05t$ for QSHE1 (Fig. \ref{figtwo} (a)) and $E=0.005t$ for
QSHE2 (Fig. \ref{figtwo} (b)). For these Fermi energies, there are
only the lowest transmission channels, i.e., the gapless edge
states. For QSHE1, the spin-up local currents locate mainly on the
lower edge (see Fig. \ref{figtwo} (a1))and the spin-down local
currents locate mainly on the upper edge (see Fig. \ref{figtwo}
(a2)). Contrary to QSHE1, for QSHE2, the spin-up local currents
locate mainly on the upper edge (see Fig. \ref{figtwo} (b1)) and
the spin-down local currents locate mainly on the lower edge (see
Fig. \ref{figtwo} (b2)). These results show that the gapless edge
states are spin-filtered and the two kinds of Rashba spin-orbit
couplings can drive an ordinary insulating state of the
silicene to the topological insulator.

\begin{figure}[htb]
\centering
\includegraphics[scale=0.45,angle=0]{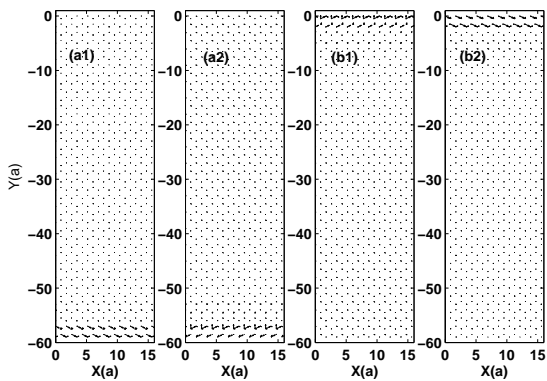}
\caption{The local spin-up (a1 and b1) and spin-down
(a2 and b2) currents density vector distribution for
$\varepsilon_{i}=0.1t$, $\lambda_{R1}=1.0t$ and $\lambda_{R2}=0.2t$
(QSHE1) with (a) $E=0.1t$ and (b) $E=-0.1t$.} \label{figthree}
\end{figure}

Next, for QSHE1, the Fermi energy is tuned to $E=0.1t$ or $E=-0.1t$,
reaching slightly into the bulk band. The configurations of
spin-dependent local-current-flow vector in such regions are plotted
in Fig. \ref{figthree}. We find that the edge states are not fully
spin-filtered when they are inside the bulk band. However, in this
case, the electrons flow along the lower edge (see Fig.
\ref{figthree} (a)), while the holes flow forward along the upper
edge (see Fig. \ref{figthree} (b)).

\begin{figure}[htb]
\centering
\includegraphics[scale=0.45,angle=0]{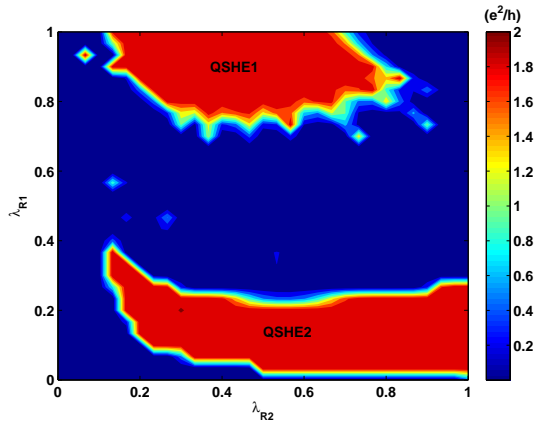}
\caption{(color online) Phase diagram showing the conductance $G$ as
a function of $\lambda_{R1}$ and $\lambda_{R2}$ for fixing
$\varepsilon_{i}=0.1t$ and $E=0$.} \label{figfour}
\end{figure}

In order to have a global view on the phase transitions, the phase
diagram at $\varepsilon_{i}=0.1t$ is plotted in Fig.
\ref{figfour}. When these two kinds of spin-orbit couplings are
tuned properly, two kinds of QSHE, QSHE1 and QSHE2 appear. For
QSHE1, the silicene nanoribbon has a large bulk gap and there are
only gapless edge states in the bulk gap because the first
spin-orbit coupling can widen the bulk gap.\cite{Ezawa1} On the
other hand, for QSHE2, the system has a narrow bulk gap and even the
gapless edge states is located in the bulk band because the second
spin-orbit coupling can narrow the bulk gap.\cite{Ezawa3}

\begin{figure}[htb]
\centering
\includegraphics[scale=0.7,angle=0]{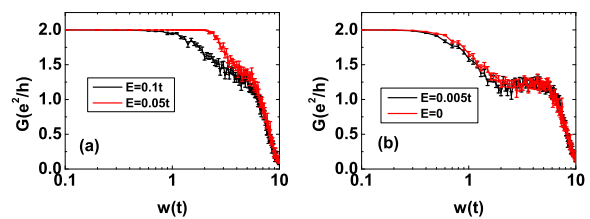}
\caption{(color online) The conductance $G$ as a function of the disorder strengths
$w$ for (a) $\varepsilon_{i}=0.1t$, $\lambda_{R1}=1.0t$ and
$\lambda_{R2}=0.2t$ (QSHE1) and (b) $\varepsilon_{i}=0.1t$,
$\lambda_{R1}=0.05t$ and $\lambda_{R2}=0.6t$ (QSHE2) with the
various energy $E$. The error bars show standard deviation of the
conductance for $100$ samples.} \label{figfive}
\end{figure}

Finally, we examine the non-magnetic disorder effect on the
conductance plateau $2e^2/h$ of the QSHE. Random on-site potential
$w_{i}$ is added for each site $i$ in the central region, where
$w_{i}$ is uniformly distributed in the range $[-w/2, w/2]$ with the
disorder strength $w$. Figs. \ref{figfive} (a) and (b) show the
conductance $G$ versus the disorder strength at various energy for
QSHE1 and QSHE2, respectively. The results show that these quantum
plateaus are robust against non-magnetic disorder because of the
topological origin of the edge states. Especially, the quantum
plateau of QSHE1 maintains its quantized value very well even when
$w$ reaches $2.0t$ for $E=0.05t$, as shown in Fig. \ref{figfive}
(a). We can also find that the gapless edge states of QSHE1 are more
insensitive to the non-magnetic disorder than those of QSHE2 because
the bulk gap of QSHE1 is larger than that of QSHE2. With further
increasing of the disorder strength, the conductance gradually
reduce to zero and the system eventually enters the insulating
regime.

In summary, we predict that the QSHE can be induced by applying an
electric field in silicene even if the intrinsic spin-orbit coupling
is very weak. The energy
bands, the configurations of the spin-dependent local-current-flow
vector, and the conductance of the system are numerically studied
using the tight-binding Hamiltonian. The first and second Rashba spin-orbit couplings,
referring to the nearest and next-nearest neighbor hoppings
respectively, can be tuned by the external electric field due to the
buckled structure of silicene. The staggered sublattice potential
induced by the external electric field open a bulk gap and the
gapless edge states are built in the gap by the two kinds of Rashba
spin-orbit couplings. With the help of spin-dependent
local-current-vector configurations, we find that the gapless edge
states are indeed spin-filtered. We also find that when the two
kinds of Rashba spin-orbit couplings are tuned properly, there are
two types of QSHE, QSHE1 with a wide bulk gap and QSHE2 with a
narrow bulk gap. Moreover, the gapless edge states have also been
found to be robust against non-magnetic disorder.

This work was supported by National Natural Science Foundation of
China (Grant Nos. 11104059, 61176089 and 11204294), Hebei province Natural
Science Foundation of China (Grant No. A2011208010), and
Postdoctoral Science Foundation of China (Grant No. 2012M510523).

\end{document}